\documentstyle[manuscript,aps]{revtex}
\begin{document}
\draft
\title{Is torsion needed in theory of gravity?}
\author{Janusz Garecki}
\address{Institute of Physics, University of Szczecin, Wielkopolska 15,
70--451 Szczecin, POLAND}
\date{\today}
\maketitle
\begin{abstract}
It is known that General Relativity ({\bf GR}) uses Lorentzian
Manifold $(M_4;g)$ as a geometrical model of the physical space--time.
$M_4$ means here a four--dimensional differentiable manifold endowed
with Lorentzian metric $g$.  The metric $g$ satisfies Einstein equations. Since
the 1970s many authors have tried to generalize this geometrical model of the
physical space--time by introducing torsion and even more general
metric--affine geometry. 
In this paper we discuss status of torsion in the theory of gravity.
At first, we emphasize that up to now we have no experimental evidence
for the existence of torsion in Nature. Contrary, the all experiments
performed in weak gravitational field (Solar System) or in strong regime
(binary pulsars) and tests of the Einstein Equivalence Principle ({\bf EEP})
confirmed {\bf GR} and Lorentzian manifold $(M_4;g)$ as correct
geometrical model of the physical space--time. 
Then, we give theoretical arguments against introducing of
torsion into geometrical model of the physical space--time.
At last, we conclude that the general--relativistic model of the physical
space--time is sufficient and it seems to be the most satisfactory.
\end{abstract}
\newpage
\section{Introduction}
In the 70ths of the XX Century many research workers have introduced torsion
into theory of gravity\footnote{We omit here older trials to introduce torsion
because they have only historical meaning.}[1,2,3]. The motivation
(only theoretical) was the following:
\begin{enumerate}
\item Analogies {\bf GR} with the continual theory of dislocations (Theory of
generalized Cosserat continuum) led to heuristic arguments for a metric
space--time with torsion, i.e., to Riemann--Cartan space--time.
\item Study of spinning matter in {\bf GR} led some people to conclusion that
the canonical energy--momentum tensor of matter $_c T_i^{~ k}$ is the source of
the curvature and the canonical intrinsic spin density tensor $_c S^{ikl} = (-)
_c S^{kil}$ is the source of torsion of the underlying space--time. From these
studies Einstein--Cartan--Sciama--Kibble ({\bf ECSK}) theory was originated and
its generalizations.
\item Trials to formulate theory of gravity as a gauge theory for Lorentz group
$L$ or for Poincare' group $P$ led to a space--time endowed with a metric--
compatible connection which could have (but not necessarily) non--vanishing
torsion, i.e., led to the Riemann--Cartan space--time [4--6].
\end{enumerate}
The above motivation {\it is not convincing}. For example, the often
used argument for torsion (It followed from 2.) based on (non-homogeneous) holonomy
theorem [1]\footnote{This theorem says that torsion gives translations and
curvature gives Lorentz rotations in tangent spaces of a
Riemann-Cartan manifold during displacements along loops.} holds only if one uses {\it Cartan
displacements} [7]. Ordinary parallell displacements gives only Lorentz
rotations (= homogeneous holonomy group) even in a Riemann-Cartan
space-time [7]. Moreover, there are other geometrical interpretation of
torsion, e.g., Bompiani [8] {\it connects torsion with rotations in tangent
spaces}, not with translations. We also needn't generalize {\bf GR} in order to get a gauge theory with $L$ or $P$ as
a gauge group [9]. Notice especially that in the {\it Asthekar's variables}
[10,11] the ordinary {\bf GR} with  Levi--Civita connection {\it itself gives} an
example of the simplest model of a gauge theory of gravity.   

Up to now, {\it we have no experimental evidence for existence of
torsion in Nature}. Contrary, the all experiments confirmed with a very high
precision ($ \sim 10^{-14}$) Einstein Equivalence Principle ({\bf EEP}) and
(with a smaller precision) General Relativity ({\bf GR}) equations [12--15]. 
Here by {\bf EEP} we mean a formulation of this Principle given by C. W.
Will [12]. In this formulation\footnote{In this formulation this
Principle can be experimentally tested.} the  {\bf EEP} states:
\begin{enumerate}
\item Weak Equivalence Principle ({\bf WEP}) is valid. 

This means that the trajectory of a freely falling body (one not acted
upon by such forces as electromagnetism and too small to be affected by
tidal forces) {\it is independent of its internal structure and
composition}. 
\item Local Lorentz Invariance ({\bf LLI}) is valid.

This means that the outcome of any local non-gravitational experiment
{\it is independent of the velocity of the freely-falling reference frame in
which it is performed}
\item Local Position Invariance ({\bf LPI}) is valid. 

This means that the outcome of any local non-gravitational experiment
{\it is independent of where and when in the Universe it is performed}.
\end{enumerate}

The only theories of gravity that can embody {\bf EEP} are those that
satisfy the postulates of {\it metric theories of gravity} [12], which
are:
\begin{enumerate}
\item Spacetime is endowed with a symmetric metric.
\item The trajectories of freely falling bodies are geodesics of that
metric.
\item In local freely falling reference frames the non-gravitational
laws of physics are those written in the language of Special Relativity
({\bf SRT}).
\end{enumerate}

From the {\bf EEP} it follows the {\it universal pure metric coupling between
matter and gravity}. This admits {\bf GR}, of course, and, at most, some of
the so--called {\it scalar--tensor theories} (these, which respect {\bf EEP}) [13--15]
without torsion. 
 
So, torsion is {\it excluded by latest gravitational experiments} (at least
in vacuum if we neglect a cosmological background\footnote{Torsion is excluded at least in vacuum because if we
neglect a cosmological background, then the all gravitational experiment
were performed in vacuum. In such a case {\bf ECSK} theory can survive
since this theory is identical in vacuum with {\bf GR}. But we think
that the arguments which we present in Section II testify against {\bf
ECSK} theory also.}) which have confirmed {\bf EEP} with such very high precision.
In consequence, torsion  can  be treated only as {\it a hypothetical field}
and is not needed in theory of gravity.

In Section II we give a review of the (theoretical) arguments which in
our opinion additionally indicate against introducing of torsion into relativistic theory of gravity. We will conclude
from these arguments that torsion rather {\it should not be introduced into
theory of gravity} .

In the paper we will confine to the metric--compatible connection
$\omega^i_{~ k}$. We cannot see any reasons in order to consider more general
connections, for example, connexion like proposed in [16]. For a metric--compatible connection $\omega^i_{~ k}$ we
have 
\begin{equation} 
Dg = 0
\end{equation}
and
\begin{equation}
\omega^i_{~ k} = _{LC} \omega^i_{~ k} + K^i_{~ k},
\end{equation}
where $_{LC} \omega^i_{~ k}$ means {\it Levi--Civita connection} and $K^i_{~
k}$ denotes {\it contortion}.

By  $\vartheta^i$ we will denote an Lorentzian coreper; $\Psi$ will
denote a matter field and $D$ will mean an exterior covariant derivative. 
For tensor fields the exterior covariant derivative $D$ reduces to an
ordinary covariant derivative $\nabla$.
\section{Arguments which indicate against torsion}
The our main argument uses Ockham's razor and the fact: a ``wonderful'',
the most simple and most symmetric Levi-Civita connection is sufficient for the all physical
requirements. 

The other our arguments are the following:
\begin{enumerate}
\item In the paper [17] the authors have showed that the geometry of
free--falls and light propagation supplemented by some (very natural axioms)
lead us to Riemannian geometry.
\item {\it Torsion is topologically trivial}. This means that the topology of a
real manifold $M$ and topology vector bundles over $M$ (determined by
characteristic classes) {\it do not depend on torsion}.  They depend {\it on
curvature only} and are determined by curvature $_{LC} \Omega^i_{~ k}$ of the
Levi-Civita connection $_{LC} \omega^i_{~ k}$ [18--20]. Roughly speaking,
one can continuously deform any metric--compatible connection (or even general
linear connection) into Levi-Civita connections {\it without changing
topological invariants and characteristic classes}. So, torsion is not relevant from
the topological point of view.
\item {\it Torsion is not relevant from the dynamical point of view too}.
Namely, one can reformulate every metric theory of gravitation with a metric-
compatible  connection $\omega^i_{~ k}$ as a "Levi-Civita theory". Torsion is
then treated as {\it a matter field}. An obvious example is given by {\bf ECSK}
theory in the so--called ``combined formulation'' [21].\footnote{In this
formulation {\bf ECSK} theory is dynamically fully equivalent to the ordinary
{\bf GR} [22].} In general, one can prove [23] that any total Lagrangian of the
type 
\begin{equation}
L_t = L_g(\vartheta^i, \omega^i_{~ k}) + L_m (\Psi, D\Psi)
\end{equation}
{\it admits an unique decomposition} into a pure geometric part ${\tilde
L}_g(\vartheta^i, _{LC} \omega^i_{~ k})$ containing no torsion plus a
generalized matter Lagrangian ${\tilde L}_m (\Psi, _{LC}D\Psi, K^i_{~ k},
_{LC}DK^i_{~ k})$ which collects the pure matter terms and all the terms
involving torsion
\begin{equation}
L_t = L_g + L_m = {\tilde L}_g + {\tilde L}_m.
\end{equation}
$_{LC}D$ means an exterior covariant derivative with respect to the
Levi-Civita connection $_{LC}\omega^i_{~k}$.
From the Lagrangian
\begin{equation}
L_t = {\tilde L}_g + {\tilde L}_m
\end{equation}
there follow the {\it Levi-Civita equations associated with} $L_t$. 

So, torsion {\it always can be treated as a matter field}. This point of view
has been taken e.g. in [24,25] and it is supported by transfromational
properties of torsion: torsion transforms like a matter field.
\item {\it Symmetry of the energy--momentum tensor of matter.} In Special
Relativity ({\bf SRT}) a correct energy--momentum tensor for matter (continuous
medium, dust, elastic body, solids) {\it must be symmetric} [26,27]. One can
always get such a tensor starting from the {\it canonical pair} $(_c T^{ik}, _c
S^{ikl}= (-) _c S^{kil})$ with properties
\begin{equation}
\partial_k {_c T^{ik}} = 0,~ ~ _c T^{ik} - _c T^{ki} = \partial_l S^{ikl},
\end{equation}
and by use {\it Belinfante symmetrization procedure} [21,28-30]
\begin{equation}
_s T^{ik} = _c T^{ik} - {1\over 2}\partial_j\bigl(_c S^{ikj} - _c S^{ijk} + _c
S^{jki}\bigr), 
\end{equation} 
\begin{equation}
S^{ijk} = _c S^{ijk} - A^{jki} + A^{ikj} = 0.
\end{equation}
Here
\begin{equation}
A^{ikj} = {1\over 2}\bigl(_c S^{ikj} - _c S^{ijk} + _c S^{jki}\bigr).
\end{equation}
The obtained new "pair" $(_sT^{ik}, ~ 0)$ is {\it the most simple and the
most symmetric}. Note that the symmetric tensor $_sT^{ik} = _s T^{ki}$
gives complete description of matter because the spin density tensor $_c S^{ijk}$
is {\it entirely absorbed} into $_s T^{ik}$ during symmetrization procedure.

It is interesting that one can easily generalize the above symmetrization
procedure onto a general metric manifold $(M_4, g)$ [9,21] by using Levi-Civita connection associated with the metric
$g$. The generalized symmetrization procedure has the same form as above 
with replacing $\eta_{ik} \longrightarrow g_{ik},~~\partial_i
\longrightarrow _{LC} \nabla_i$. 

So, one can always get in a metric manifold $(M_4,g)$ a symmetric
energy--momentum tensor $_sT^{ik} = _sT^{ki}$ for matter (then, of course, corresponding
$S^{ikj} = 0$). Observe that the symmetric tensor $_sT^{ik}$, like as in {\bf SRT},
consists of $_cT^{ik}$ and $_cS^{ikl}$.

The symmetric energy--momentum tensor for matter is uniquely determined by the
matter equations of motion [31]. This fact is very important for the uniqueness
of the gravitational field equations. Moreover, the symetric energy--momentum
tensor is covariantly conserved.

L. Rosenfeld has proved [32] that 
\begin{equation}
_s T^{ik} = {\delta L_m\over\delta g_{ik}},
\end{equation}
where $L_m = L_m(\Psi,~ _{LC}D\Psi)$ is a covariant Lagrangian density for
matter.

The tensor $_s T^{ik}$ given by (10) is the source in the Einstein
equations 
\begin{equation}
G_{ik} = \chi _s T_{ik},
\end{equation}
where $\chi = {8\pi G\over c^4}$.

Note that these equations {\it geometrize the both canonical quantities}
$_c T^{ik}$ and $_c S^{ikl} = (-)_c S^{kil}$ {\it in a some equivalent
way} because  the tensor $_s T^{ik}$ is built from these two canonical
tensors. 

So, it is the most natural and most simple to postulate that, in general, the
correct energy--momentum tensor for matter is the symmetric tensor $_sT^{ik}$.
This leads us to a pure metric, torsion-free, theory of gravity which has the
field equations of the form
\begin{equation}
{\delta L_g\over\delta g_{ik}} = {\delta L_m\over\delta g_{ik}}.
\end{equation}
Then, if we take into account the {\it universality of the Einstein equations}
[33--38], we will end up with General Relativity equations
(possible with $\Lambda \not= 0$) which will have a sophisticated,
symmetric energy-momentum tensor as a source.

\item A gravitational theory with torsion violates {\bf EEP} which has
so very good experimental evidence (up to $10^{-14}$).
\item In a space-time with torsion a tangent space $T(P)$ cannot be
identified with Minkowskian spacetime because there do not exist holonomic coordinates in
which $g_{ik}(P) = \eta_{ik}, ~~\Gamma^i_{kl}(P) = 0$. So, a gravitational
theory with torsion {\it is not a covering theory for} {\bf SRT}. We
also lose Fermi coordinates [7,39] in Riemann-Cartan
space-time.\footnote{ Fermi coordinates realize in {\bf GR} a local
(freely-falling) inertial frame in which {\bf SRT} is valid.}
\item A low-energetic superstrings gravity needn't torsion. It uses only
metric $g$ and scalar fields and can always be formulated as Einstein
theory (in ``Einstein frame'') without torsion.
\item A connection having torsion can be determined neither by its own
autoparallells (paths) nor by geodesics [7]. So, one cannot determine a
connection which has torsion by observation of the test particles (which
move along geodesics or autoparallells).
\end{enumerate}

Torsion leads to ambiguities:
\begin{enumerate}
\item {\it Minimal Coupling Principle} ({\bf MCP}) $\not=$ {\it Minimal
Action Principle} ({\bf MAP}) in a space-time with torsion [29].

{\bf MCP} can be formulated as follows. In a {\bf SRT} field equations
obtained from the {\bf SRT} Lagrangian density $L = L(\Psi ,
\partial_i\Psi)$ we replace
$\partial_i\longrightarrow\nabla_i,~~\eta_{ik}\longrightarrow g_{ik}$
and get {\it covariant field equations} in $(M_4,g)$.

By {\bf MAP} we mean an application of the Minimal Action Principle
(Hamiltonian Principle) to the covariant action integral $S =
\int\limits_{\Omega} L(\Psi,~\nabla_i\Psi)d^4\Omega$, where $L(\Psi,
~\nabla_i\Psi)$ is a covariant Lagrangian density obtained from the {\bf
SRT} Lagrangian density $L(\Psi,~\partial_i\Psi)$ by {\bf MCP}.

It is natural to expect that the field equations in $(M_4,g)$ obtained by
using {\bf MCP} on {\bf SRT} equations should coincide with the
Euler-Lagrange equations obtained from $L(\Psi,~\nabla_i\Psi)$ by {\bf
MAP}. {\it This holds in GR but not in the framework of the
Riemann-Cartan geometry}. So, we have there an ambiguity in the field equations.\footnote{Axial torsion
removes this ambiguity. By $(M_4,g)$ we mean here a general metric
manifold; not necessarily Riemannian.} 
\item In the framework of the {\bf ECSK} theory of gravity we have four
energy-momentum tensors for matter: Hilbert, canonical, combined, formal
[21]. Which one is correct?
\item Let us consider now normal coordinates {\bf NC(P)} [7,40] which
are so very important in {\bf GR} (See, eg., [41-43]). In the framework of the Riemann-Cartan geometry we have
two {\bf NC(P)}: normal coordinates for the Levi-Civita part of the
Riemann-Cartan connection ${\bf NC(_{LC}\omega,~P)}$ and normal
coordinates for the symmetric part of the full connection ${\bf
NC(_s\omega,~P)}$ [44,45]. Which one has a greater physical meaning? 

The above ambiguity of the normal coordinates\footnote{Axial torsion
removes this ambiguity.} leads us to ambiguities in
superenergy and supermomentum tensors [44,45]. Moreover, the obtained
expressions are too complicated for practical using. In fact, we lose
here a possibility of efective using of the normal
coordinates.
\item In the framework of the Riemann-Cartan geometry [7]
\begin{equation}
R_{(ik)lm} = R_{ik(lm)} = 0,
\end{equation}
but
\begin{equation}
R_{iklm}\not= R_{lmik}.
\end{equation}
The last asymmetry leads to an ambiguity in construction of the
so--called ``Maxwellian superenergy tensor'' for the field $R_{iklm}$
[46]. This tensor is uniquely constructed in {\bf GR} owing to the
symmetry $R_{iklm} = R_{lmik}$ and it is proportional to the
Bel-Robinson tensor [47,48]. In the framework of the Riemann-Cartan
geometry the obtained result depends on which antysymmetric pair of the
$R_{iklm}$, first or second, is used in construction. 
\item In a Riemann-Cartan space-time we have geodesics and
autoparallells (paths). Hamiltonian Principle demands geodesics as
trajectories for the test particles. Then, what about physical meaning of
the autoparallells? \footnote{Axial torsion removes this problem.}
\item In a space-time with torsion we have in fact two kinds of
parallell displacement defined by
\begin{equation}
dv^k = (-)\Gamma^k_{ij}v^j dx^i
\end{equation}
and
\begin{equation}
dv^k = (-)\Gamma^k_{ij}v^i dx^j.
\end{equation}
There follow from that two kinds of absolute (and covariant) differentials 
\begin{equation}
\nabla_i v^k = \partial_i v^k + \Gamma^k_{il}v^l
\end{equation}
\begin{equation}
\nabla_i v^k = \partial_i v^k + \Gamma^k_{li}v^l.
\end{equation}
Which one of the two above possibilities could be eventually realized in
Nature? 

In practice, one must consequently use one of the two above
possibilities (or conventions) during any calculations in order to avoid
mistakes. For example, in Hehl's papers 
\begin{equation}
\nabla_i\sqrt{\vert g\vert} = 0
\end{equation}
in the Riemann-Cartan space-time (and it is, of course, correct result).
But in the paper [49] you can find 
\begin{equation}
\nabla_i\sqrt{\vert g\vert} \not= 0.
\end{equation}
The last result is, of course, uncorrect and it is a consequence of
mixing of the two above covariant differentiation.
\end{enumerate}

The other source of the computational mistakes connected with torsion is
the following: different Authors use definitions of torsion which differ
by sign and by factor $1/2$.
\section{Concluding remarks}
As we have seen, the {\bf GR} model of the space-time has very good
experimental confirmation. On the other hand, torsion has no
experimental evidence (at least in vacuum) and it is not needed in the
theory of gravity. Moreover, introducing of torsion into geometric
structure of the space-time leads us to many problems (apart
from calculational, of course). Some of these problems removes an axial
torsion $A_i = {1\over 6}\eta_{iabc}Q^{abc}, ~Q^{[abc]} = Q^{abc}$. So,
it would be reasonable to confine themselves to the axial torsion only
(If one still want to keep on torsion). This is also supported by an
important fact that the matter fields (= Dirac's particles) are coupled
only to the axial part of torsion in the Riemann-Cartan space-time.
However, if we confine to the axial torsion, then (if you remember
dynamical triviality of torsion and universality of the Einstein equations) we effectively
will end up with {\bf GR} + an additional pseudovector field $A_i$ (or with
an additional pseudoscalar field $\varphi$ if the field $A_i$ is
potential, i.e., if $A_i = \partial_i\varphi$) [29]. But {\bf GR} with
an additional dynamical pseudovector field $A_i$ yields local
gravitational physics which may have both location and
velocity-dependent effects [13] unobserved up to now. So, we will finish
with the conclusion that the geometric model of the space-time given by
ordinary {\bf GR} with ``wonderful'' Levi-Civita connection {\it seems to be the
most satisfactory}. \footnote{During Symposium LISA 3 (Golm, July 10-14,
2000) I have discussed with many Participiants of the Symposium about
geometric model of the space-time. They all agreed with this
conclusion. See also conclusion about correct theory of gravity and
space-time structure given recently in [50].} 
\centerline{\bf Acknowledgments}

I would like to thank Prof. F.W. Hehl and Prof. A. Trautman for their,
most useful, critical remarks. I would like also to emphasize that I have learned
at first about possible physical meaning of torsion from excellent papers
given by them years ago.  Under influence of these papers I was a
fanatic follower of torsion in past. But recently, under influence of
the papers on experimental gravity which are cited in References, papers
on universality of the Einstein equations and argumentation given in
Section II, I have changed my mind.

\end{document}